\documentclass{aastex6}

\renewcommand\thefootnote{\arabic{footnote}}

\begin{document}


\title{Cosmic ray short burst observed with the Global Muon Detector Network (GMDN) on June 22, 2015}

%% Use \author, \affil, and the \and command to format author and affiliation 
%% information.  If done correctly the peer review system will be able to
%% automatically put the author and affiliation information from the manuscript
%% and save the corresponding author the trouble of entering it by hand.
%%
%% The \affil should be used to document primary affiliations and the
%% \altaffil should be used for secondary affiliations, titles, or email.

%% Authors with the same affiliation can be grouped in a single
%\author{Greg J. Schwarz\altaffilmark{1,2} and August Muench\altaffilmark{1}}
%\affil{American Astronomical Society \\
%2000 Florida Ave., NW, Suite 300 \\
%Washington, DC 20009-1231, USA}

%\author{Butler Burton\altaffilmark{3}}
%\affil{National Radio Astronomy Observatory}

%\affil{TeXnology Inc}

%\author{Julie Steffen\altaffilmark{4}}
%\affil{American Astronomical Society \\
%2000 Florida Ave., NW, Suite 300 \\
%Washington, DC 20009-1231, USA}

%% Use the \and command so offset the last author.
%\and

\author{K.~Munakata\altaffilmark{1}, M.~Kozai\altaffilmark{2}, P.~Evenson\altaffilmark{3}, T.~Kuwabara\altaffilmark{3}, C.~Kato\altaffilmark{1}, M.~Tokumaru\altaffilmark{4}, M. Rockenbach\altaffilmark{5}, A. Dal Lago\altaffilmark{5}, R. R. S. Mendonca\altaffilmark{5,6}, C. R. Braga\altaffilmark{5}, N. J. Schuch\altaffilmark{7},  H. K. Al Jassar\altaffilmark{8}, M. M. Sharma\altaffilmark{8}, M. L. Duldig\altaffilmark{9}, J. E. Humble\altaffilmark{9}, I. Sabbah\altaffilmark{10}, and J. K\'ota\altaffilmark{11}}

\altaffiltext{1}{Department of Physics, Shinshu University, Matsumoto, Nagano 390-8621, Japan}
\altaffiltext{2}{Institute of Space and astronomical Science, Japan Aerospace Exploration Agency (ISAS/JAXA), Sagamihara, Kanagawa 252-5210, Japan}
\altaffiltext{3}{Bartol Research Institute and Department of Physics and Astronomy, University of Delaware, Newark, DE 19716, USA}
\altaffiltext{4}{Institute for Space-Earth Environmental Research, Nagoya University, Nagoya, Aichi 464-8601, Japan}
\altaffiltext{5}{National Institute for Space Research (INPE), 12227-010 S\~{a}o Jos\'{e} dos Campos, SP, Brazil}
\altaffiltext{6}{State Key Laboratory of Space Weather, National Space Science Center (NSSC), Chinese Academy of Sciences, NO. 1 Nanertiao, Zhongguancun, Beijing, 100190, China}
\altaffiltext{7}{Southern Regional Space Research Center (CRS/INPE), P.O. Box 5021, 97110-970, Santa Maria, RS, Brazil}
\altaffiltext{8}{Physics Department, Kuwait University, P.O. Box 5969 Safat, Kuwait 13060}
\altaffiltext{9}{School of Natural Sciences, University of Tasmania, Hobart, Tasmania 7001, Australia}
\altaffiltext{10}{Department of Natural Sciences, College of Health Sciences, Public Authority of Applied Education and Training, Kuwait City 72853, Kuwait}
\altaffiltext{11}{Lunar and Planetary Laboratory, University of Arizona, Tucson, AZ 87721, USA}

\begin{abstract}
We analyze the short cosmic ray intensity increase (``cosmic ray burst'': CRB) on June 22, 2015 utilizing a global network of muon detectors and derive the global anisotropy of cosmic ray intensity and the density (i.e. the omnidirectional intensity) with 10-minute time resolution. We find that the CRB was caused by a local density maximum and an enhanced anisotropy of cosmic rays both of which appeared in association with Earth's crossing of the heliospheric current sheet (HCS). This enhanced anisotropy was normal to the HCS and consistent with a diamagnetic drift arising from the spatial gradient of cosmic ray density, which indicates that cosmic rays were drifting along the HCS from the north of Earth. We also find a significant anisotropy along the HCS, lasting a few hours after the HCS crossing, indicating that cosmic rays penetrated into the inner heliosphere along the HCS. Based on the latest geomagnetic field model, we quantitatively evaluate the reduction of the geomagnetic cut-off rigidity and the variation of the asymptotic viewing direction of cosmic rays due to a major geomagnetic storm which occurred during the CRB and conclude that the CRB is not caused by the geomagnetic storm, but by a rapid change in the cosmic ray anisotropy and density outside the magnetosphere.
\end{abstract}

\keywords{cosmic rays --- interplanetary medium --- Sun: coronal mass ejections (CMEs)}



\section{Introduction} \label{sec:intro}
%CME & Space weather

The GRAPES-3 muon telescope observed a “cosmic ray burst” (CRB) in which the muon count rate increased  $\sim$1~\% for two hours during June 22, 2015 \citep{Mohanty16} [hereafter referred to as Paper 1, see also \citet{Non06} for GRAPES-3]. This burst was recorded shortly after the arrival of a strong interplanetary shock identified by abrupt increases of the solar wind velocity and the interplanetary magnetic field (IMF) strength. There were two preceding shocks recorded about 15 and 27 hours earlier. These three shocks were formed in front of coronal mass ejections (CMEs) successively ejected from the same active region NOAA 2371. The third shock was followed by a strong enhancement of south directing IMF, which triggered a major geomagnetic storm reaching a $\it{K_{p}}$ index of 8+. Paper 1 proposed that the CRB was caused by the substantial reduction of the geomagnetic cut-off rigidity and the alteration of asymptotic cosmic ray orbits in the magnetosphere.\par

In this paper, we analyze the CRB observed with the Global Muon Detector Network (GMDN) comprising four multidirectional muon detectors located in Japan, Australia, Brazil and Kuwait, designed for accurate observation of the global cosmic ray anisotropy. In our previous analyses of the GMDN data, the cosmic ray anisotropy and density have been derived on an hourly basis and used to analyze the geometry of the cosmic ray depletion region in individual Forbush decreases \citep{Kuw09,Koz16}. Here, we use 10-minute data to resolve and analyze the rapidly changing anisotropy and density in the CRB to show that the CRB was caused by a rapid enhancement of the cosmic ray anisotropy and density outside the magnetosphere, and not by the geomagnetic storm.\par

We describe the GMDN and data analysis in Sections \ref{subsec:gmdn} and \ref{subsec:analysis}, respectively, and show the results in Section \ref{subsec:results}. We give a summary and discussion in Section \ref{sec:discuss}.\par

\section{Data Analysis and results} \label{sec:exp}

\subsection{Global Muon Detector Network (GMDN)} \label{subsec:gmdn}

The GMDN comprises four multidirectional muon detectors, ``Nagoya'' in Japan, ``Hobart'' in Australia, ``Kuwait City'' in Kuwait and ``S\~ao  Martinho da Serra'' in Brazil, recording muon count rates in 60 directional channels viewing almost the entire sky around Earth. Each detector except Kuwait City consists of two horizontal layers separated by 1.73~m of 1~m$^{2}$ plastic scintillators (PSs), each viewed by a 12.7~cm diameter photomultiplier tube. By counting twofold coincidences between pairs of detectors in the upper and lower layers, we record the rate of muons from the corresponding incident direction. Kuwait City consists of four horizontal layers of proportional counter tubes (PCTs), each 5~m long with a 10~cm diameter with a 50~$\mu$m thick tungsten anode along the cylinder axis. The PCT axes are aligned east-west (X) in the top and third layers and north-south (Y) in the second and bottom layers. The top and second layers form an upper pair, while the third and bottom layers form a lower pair. The two pairs are separated vertically by 80~cm. Muon recording is triggered by the fourfold coincidence of pulses from all layers and the incident direction is identified from X-Y locations of the upper and lower PCT pairs.\par

Table~\ref{tab1} summarizes characteristics of directional channels of the GMDN, while Figure~\ref{map} shows the asymptotic viewing directions (corrected for geomagnetic bending of cosmic ray orbits) of 60 directional channels of the GMDN. The detector configurations in the GMDN are also available at our web-site\footnote{https://cosray.shinshu-u.ac.jp/crest/DB/Documents/documents.php}. In our calculations of the median primary rigidity ($P_{m}$) and the asymptotic viewing direction ($\phi_{\rm asymp}$ \& $\lambda_{\rm asymp}$) at $P_{m}$ of each directional channel, we use the response function of the atmospheric muons to the primary cosmic rays given by numerical solutions of the hadronic cascade in the atmosphere \citep{Mur79}. We also performed Monte Carlo (MC) simulations of the hadronic cascade by using CORSIKA (HDPM+GHEISHA), calculated the response function and found that both response functions are in a good agreement except for minor differences in the high-energy region above 1~TeV. The difference between the calculated median primary rigidities is less than $\sim$5~\%  even for the most inclined directional channels. We also find that the observed count rates are $\sim$5 to 15~\% higher than the calculations. This $\sim$15~\% underestimation of our calculations, however, is expected to be constant and does not affect our analyses in this paper based on the fractional intensity (see Section \ref{subsec:analysis}).\par

A 5 cm-thick layer of lead is installed in each detector to absorb the soft component radiation in the air. The muon threshold energy is 300~MeV for the vertical directional channels and 1300~MeV for the most inclined directional channels. In our detectors, coincidence between any pair of layers triggers a muon count. If there are multi-hits on a layer due to delta-rays produced by muons traversing the 5~cm lead layer, the total count in directional channels exceeds the total count of muons. To evaluate this effect, we performed MC simulations for the Nagoya muon detector in which each layer consists of a 6$\times$6 horizontal array of 1~m$^{2}$ plastic scintillators. We used GEANT4 (version 10.3 patch-03, 20-October-2017) for our simulations. We randomly generated muons with various energies and incident directions, produced hit-patterns in each layer of 1~m$^{2}$ plastic scintillators and calculated counts in each directional channel. It was found that the ratio of the multi-hit event number to the trigger number increases with increasing muon energy and zenith angle, while the muon flux at Nagoya decreases. By calculating the ratio expected for 17 directional channels available in Nagoya as a function of monitored zenith angle, we found that the contribution from the multi-hit events to the total count rate is about 3~\% in the vertical channel and about 15~\% in the most inclined directional channels. Delta-rays affect the absolute count rate, but the contribution is expected to be constant and does not affect our analyses in this paper based on the fractional intensity. The contribution of delta-rays also slightly increases the median primary energy, but the difference from the value in Table \ref{tab1} is less than a few GV at most and very small when the broad energy response is considered.\par

\subsection{Data analysis} \label{subsec:analysis}

We analyze each percent deviation of the pressure corrected 10-minute muon count rate, $I_{i,j}(t)$ in the $j$-th directional channel of the $i$-th detector in the GMDN at universal time $t$, from the monthly mean in June, 2015. One minute count rates are also available from Nagoya, Hobart and S\~ao  Martinho da Serra, but not from Kuwait City which was being enlarged in 2015. After the enlargement of Kuwait City completed in March 2016, one minute data have been available from all detectors.\par

We model $I_{i,j}(t)$ in terms of the cosmic ray density (or omnidirectional intensity) $I_0(t)$ and three components $\left( \xi^{\rm GEO}_{x}(t), \xi^{\rm GEO}_{y}(t), \xi^{\rm GEO}_{z}(t) \right)$ of the first order anisotropy vector ({\boldmath{$\xi$}}$^{\rm GEO}(t)$) in the geographic (GEO) coordinate system, as
\begin{eqnarray}
   I^{fit}_{i,j}(t) = I_0(t)c_{0 i,j}^0 &+& \xi^{\rm GEO}_x(t)(c_{1 i,j}^1 \cos \omega t_i - s_{1 i,j}^1 \sin \omega t_i) \nonumber \\
   &+& \xi^{\rm GEO}_y(t)(s_{1 i,j}^1 \cos \omega t_i + c_{1 i,j}^1 \sin \omega t_i) \nonumber \\
   &+& \xi^{\rm GEO}_z(t) c_{1 i,j}^0
   \label{eq:fit}
\end{eqnarray}
where $t_{i}$ is the local time in hours at the $i$-th detector, $c^{0}_{0 i,j}$, $c^{1}_{1 i,j}$, $s^{1}_{1 i,j}$ and $c^{0}_{1 i,j}$ are coupling coefficients and $\omega=\pi/12$. In the GEO coordinate system, we set the $x$-axis to the anti-Sun direction in the equatorial plane, the $z$-axis to the geographical north perpendicular to the equatorial plane and the $y$-axis completing the right-handed coordinate system. The coupling coefficients are calculated using the response function of the atmospheric muon intensity to primary cosmic rays mentioned above \citep{Nag71,Fuj84}. We derive the best-fit set of four parameters $\left( I_0(t), \xi^{\rm GEO}_{x}(t), \xi^{\rm GEO}_{y}(t), \xi^{\rm GEO}_{z}(t) \right)$ by solving the following linear equations.
\begin{eqnarray}
\frac{\partial S}{\partial I_0(t)}=\frac{\partial S}{\partial \xi^{\rm GEO}_{x}(t)}=\frac{\partial S}{\partial \xi^{\rm GEO}_{y}(t)}=\frac{\partial S}{\partial \xi^{\rm GEO}_{z}(t)}=0
   \label{eq:lsm}
\end{eqnarray}
where $S$ is the residual defined, as
\begin{eqnarray}
S=\sum_{i,j}{(I_{i,j}(t)-I^{fit}_{i,j}(t))^{2}/\sigma_{ci,j}^2}
   \label{eq:res}
\end{eqnarray}
and $\sigma_{ci,j}$ is the count rate error of $I_{i,j}(t)$.

The data analysis method based on this equation has been shown to be valid, useful and successful in several previous papers \citep{Kuw04,Mun05,Oka08,Fus10,Rock11,Rock14,Koz14,Koz16}. The derived anisotropy vector {\boldmath{$\xi$}}$^{\rm GEO}(t)$ in the GEO coordinate system is then transformed to {\boldmath{$\xi$}}$^{\rm GSE}(t)$ in the geocentric solar ecliptic (GSE) coordinate system for comparisons with the solar wind and IMF data. As seen in Equation \ref{eq:fit}, the observed $I_{i,j}(t)$ varies depending on both $I_0(t)$ and {\boldmath{$\xi$}}$^{\rm GEO}(t)$ which must be determined separately from $I_{i,j}(t)$. Although the muon count rate in each detector of the GMDN is much smaller than that in GRAPES-3, the global sky-coverage of the GMDN seen in Figure~\ref{map} makes it possible to determine the anisotropy and density, separately and accurately. We will discuss this in Section \ref{sec:discuss}.\par

\subsection{Results} \label{subsec:results}

\begin{figure}[ht!]
\epsscale{1.0}
\plotone{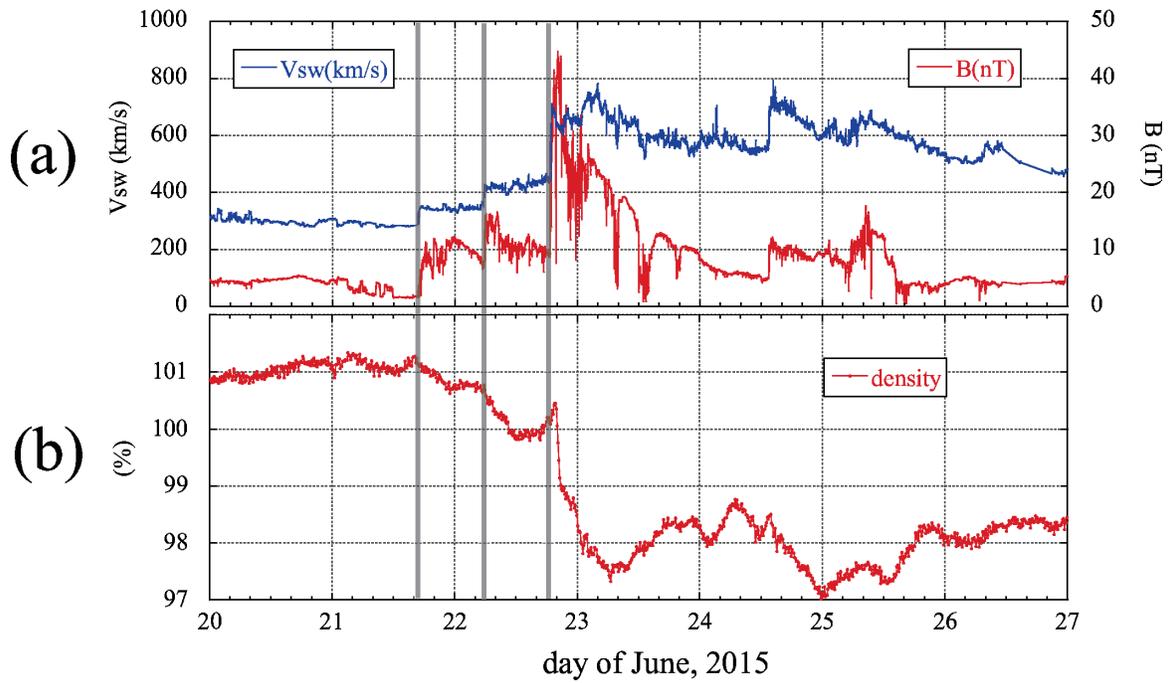}
\caption{Solar wind parameters and cosmic ray density between June 20 and 26, 2015. Panel (a) displays 10-minute averages of solar wind velocity (blue curve) and IMF strength (red curve), while panel (b) shows the cosmic ray density derived from best-fitting to 10-minute GMDN data. Arrival times of three shocks are indicated by gray vertical lines.
\label{den}}
\end{figure}

Figure~\ref{den}(b) displays the best-fit density ($I_0(t)$) during seven days including the CRB on June 22, 2015, together with 10-minute averages of the solar wind velocity and the IMF strength in Figure~\ref{den}(a). It is seen that $I_0(t)$ decreases responding to each arrival of three shocks indicated by vertical gray lines, while it starts increasing in the second half of June 22 before the third shock arrival toward the local maximum in the same day and rapidly decreases afterward. The local maximum of $I_0(t)$ in June 22 is lower than the level before the second shock arrival, but higher than the level before the third shock arrival. As will be shown later, this implies that cosmic rays forming the local maximum are transported from the region between the second and third shocks where the cosmic ray intensity is less depleted by the second shock than at Earth. We note that the MHD simulation of the space weather actually reproduces such a region being formed by the third shock overtaking the second shock, which extends to the north of Earth when the third shock arrived at Earth\footnote{see archived data enlil\_com1\_20150622T220000 available at https://www.ngdc.noaa.gov/enlil/}.\par

\begin{figure}[ht!]
\epsscale{1.0}
\plotone{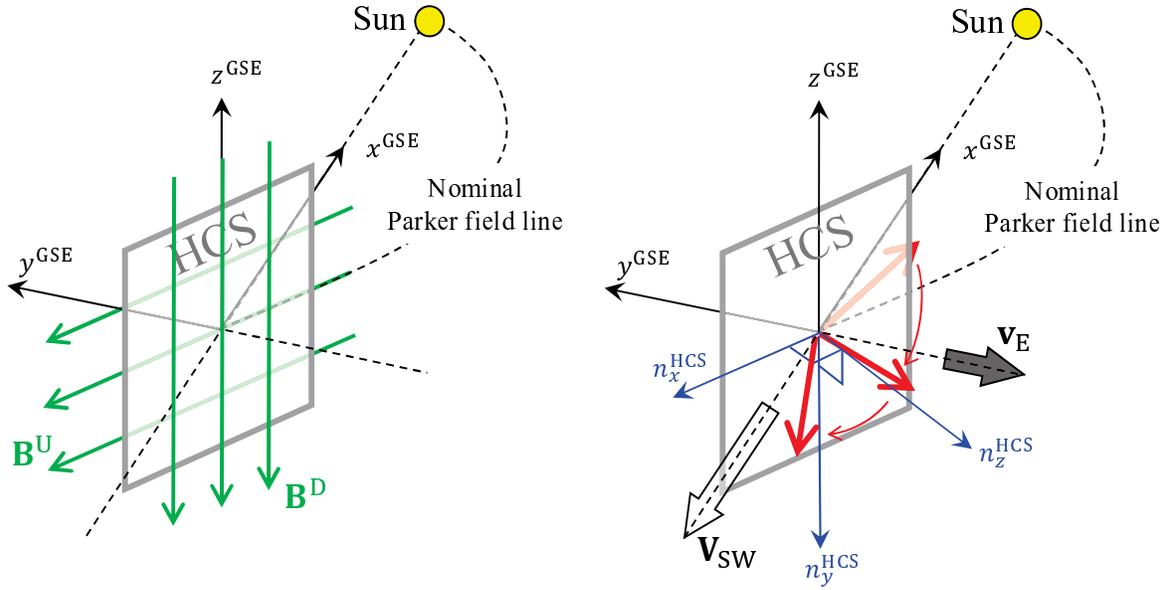}
\caption{Illustration of the IMF and anisotropy on both sides of the HCS. The left panel displays the up-wind and down-wind IMF vectors ($\bf{B}^{\rm{U}}$ and $\bf{B}^{\rm{D}}$) by green vectors on both sides of the HCS indicated by a plane with gray frames, while the right panel illustrates the solar wind velocity ($\bf{V}_{\rm{SW}}$) and the velocity of Earth's orbital motion ($\bf{v}_{\rm{E}}$), which are used for the corrections of the solar wind convection and Compton-Getting effect, by unfilled and gray filled arrows, respectively. Black axes on both panels represent the GSE coordinate system, while blue axes on the right panel represent the HCS coordinate system in which z-axis directs parallel to the normal vector of the HCS. Three red arrows in the right panel illustrate the observed cosmic ray streaming orientation (opposite to the anisotropy) rapidly changing across the HCS (see Figure~\ref{ani}c and text).
\label{sch}}
\end{figure}

Figure~\ref{obs}(a)-(c) display 1-minute solar wind data from the OMNIWeb dataset during the second half of June 22 including the CRB\footnote{https://omniweb.gsfc.nasa.gov}, while solid circles in panels (d)-(g) show $I_{i,j}(t)$ recorded in four vertical channels of the GMDN for the same period. Following abrupt increases in the solar wind velocity and IMF strength at 18:39 UT in Figure~\ref{obs}(a), indicated by a gray vertical line, the strong southward IMF ($B_{z}<0$ shown by red curve in Figure~\ref{obs}(c)) discontinuously reduces, while $B_{x} (<0)$ and $B_{y} (>0)$ (shown by blue and green curves, respectively) become significant with similar magnitudes at 19:46 indicated by the vertical dotted line. This indicates Earth's crossing of the tangential discontinuity or the Heliospheric Current Sheet (HCS) at this time and the southward IMF changing its orientation to almost parallel to the nominal Parker field ($B_{lat}\sim$0\degr, $B_{long}\sim$135\degr) directed away from the Sun along the Archimedean line. By using 10-minute averages of the IMF, we calculate the normal vector ($\bf{n}^{\rm{HCS}}$) to this HCS from the vector product ($\bf{B}^{\rm{U}}\times\bf{B}^{\rm{D}}$) between the observed southward IMF $\bf{B}^{\rm{D}}$ ($B^{D}_{lat}$=-74.3\degr, $B^{D}_{long}$=327.5\degr) in the down-wind direction and the IMF $\bf{B}^{\rm{U}}$ ($B^{U}_{lat}$=-0.5\degr, $B^{U}_{long}$=133.3\degr) in the up-wind direction \citep{BurNes69}. The GSE-latitude and longitude of the calculated normal vector are  -3.7\degr and 227.1\degr, respectively, indicating that the HCS at Earth was nearly perpendicular to the ecliptic plane. Figure~\ref{sch} illustrates the geometries of the IMF and HCS.\par

The muon count rates in Figures~\ref{obs}(d)-(f) show local maxima around this HCS at similar times to the CRB reported by Paper 1, except the rate of Brazilian detector in Figure~\ref{obs}(g), which views in almost the opposite direction to Nagoya-V in Figure~\ref{obs}(d) (see Figure~\ref{map}). This indicates a significant contribution from the global anisotropy to the CRB. The dotted curves in Figures~\ref{obs}(d)-(g) show $I^{fit}_{i,j}(t)$ reproduced using the best-fit parameters in Equation \ref{eq:fit}, while the red and blue curves represent contributions to $I^{fit}_{i,j}(t)$ from the cosmic ray anisotropy ({\boldmath{$\xi$}}$^{\rm GEO}(t)$) and density ($I_0(t)$), respectively. It is clear that the contribution from the anisotropy (red curves) is out-of-phase for Nagoya in Japan and S\~ao  Martinho da Serra in Brazil, enhancing (canceling) the common increase due to the density (blue curves) in Nagoya (S\~ao  Martinho da Serra).\par

Figure~\ref{ani} shows the best-fit cosmic ray density and anisotropy observed in every 10 minutes, over the same period as Figure~\ref{obs}. The shock arrival and the HCS crossing are again indicated by vertical gray and dotted lines, respectively. The error of each parameter is deduced from the muon count rate error ($\sigma_{ci,j}$) (see Figure~\ref{err} in Section \ref{sec:discuss}) and the dispersion of 1-minute values of each IMF parameter. It is seen in Figure~\ref{ani}(a) that the cosmic ray density ($I_0(t)$) gradually increases toward a local maximum just at the HCS crossing and rapidly decreases afterward. Due to this local maximum, the decrease of $I_0(t)$ appears to start about one hour after the shock arrival. The amplitude of the anisotropy ({\boldmath{$\xi$}}$^{\rm GSE}(t)$) displayed by the green curve in Figure~\ref{ani}(b) also shows a local maximum around HCS and remains large at $\sim$~1~\% afterward. It is clear that the anisotropy component perpendicular to the IMF (red circles) dominate the total anisotropy (green curve) around the HCS crossing. A striking feature is the anisotropy orientation changing systematically around the HCS in Figure~\ref{ani}(c). This implies that an anisotropy vector with a significant amplitude rapidly passed the field of view (FOV) of GMDN detectors and was observed as CRBs in some directional channels.\par

Shown in Figure~\ref{ani}(d) is the anisotropy after subtracting the solar wind convection and the Compton-Getting anisotropy arising from Earth's orbit around the Sun \citep{Ame04}, calculated by using 10-minute average of the solar wind velocity $\bf{V}_{\rm{SW}}$ (blue curve in Figure~\ref{obs}(a)) and the velocity of Earth's orbital motion $\bf{v}_{\rm{E}}$ of 30~km/s opposite to the GSE-y axis (see Figure~\ref{sch}). These corrections are made by adding a vector $(2+\gamma)[\bf{V}_{\rm{SW}}-\bf{v}_{\rm{E}}]/\it{c}$ to {\boldmath{$\xi$}}$^{\rm GSE}(t)$, where $\it{c}$ is the speed of light and $\gamma$ is the power-law index of the GCR energy spectrum set to 2.7 (e.g. \citet{Oka08}). This corrected anisotropy is solely due to the diffusion and the diamagnetic drift which reflect the spatial distribution of cosmic rays. Seen more clearly in this figure than in Figure~\ref{ani}(b) is a local enhancement of the anisotropy (green curve) around the HCS. This enhanced anisotropy causes the rapid change in the anisotropy orientation seen in Figure~\ref{ani}(c).\par

The duration of the rapid change of the anisotropy is about two hours and comparable to the time scale ($R_{L}/V_{SW}\sim$~2.7~hours) for the solar wind to travel across the Larmor radius ($R_{L}\sim$~0.04~AU) of 60~GV cosmic rays in IMF ($B\sim$~35~nT) with the average velocity ($V_{SW}\sim$~650~km/s) in the period between 19:00 and 21:00 UT. This implies that the anisotropy and spatial distribution of cosmic rays with large $R_{L}$ varies only gradually, rather than instantaneously responding to an abrupt change in the IMF orientation across the HCS.\par

As seen in Figure~\ref{ani}(d), the local enhancement of the anisotropy is dominated by the component perpendicular to the IMF (red circles). By ignoring the contribution from the perpendicular diffusion, i.e.~assuming that the perpendicular anisotropy is solely arising from diamagnetic drift, we deduce the spatial density gradient vector ($\bf{G}$) perpendicular to the IMF by using 10-minute average of the observed IMF in Figure~\ref{obs}. Figure~\ref{ani}(e) shows the three GSE components of $\bf{G}$. The positive $G_{z}$ (red circles) increases following the HCS crossing indicating the ``source'', from which cosmic rays are transported, probably by the drift along HCS, is located north of Earth. It is noted here that the  drift transporting cosmic rays may alter the preexisting spatial distribution of cosmic rays, but it cannot be directly observed as a directional anisotropy along the HCS. Instead, the spatial distribution of cosmic rays produced by the drift is observed by the diamagnetic drift anisotropy which is perpendicular to the HCS.\par

Figure~\ref{ani}(f) shows the anisotropy components parallel and perpendicular to the HCS calculated by using the HCS normal vector ($\bf{n}^{\rm{HCS}}$) defined above. The parallel component (blue circles) dominates the anisotropy after the local enhancement of the anisotropy, while the perpendicular component (red circle) becomes dominant around the HCS. We calculate the anisotropy orientation in a coordinate system fixed to the HCS in which the z-axis is parallel to $\bf{n}^{\rm{HCS}}$, the x-axis is parallel to the up-wind IMF ($\bf{B}^{\rm{U}}$) and the y-axis completes the right-handed coordinate system (see blue axes on the right panel of Figure~\ref{sch}). The calculated longitude and latitude of the anisotropy vector in this coordinate system are shown in Figure~\ref{ani}(g). The anisotropy latitude and longitude are both around zero after the HCS crossing, indicating that the anisotropy is consistent with the streaming along the HCS, nevertheless the observed local IMF orientation shows rapid and large variations  (see Figure~\ref{obs}(c)).\par

\section{Summary and Discussions} \label{sec:discuss}

We analyzed the CRB observed by the GMDN on June 22, 2015 and found that the CRB was caused by the enhanced cosmic ray anisotropy and density outside the geomagnetic field around the HCS, where the southward IMF changed its orientation to the nominal Parker field. While the IMF orientation changes almost instantaneously at the HCS, the anisotropy of 60~GV cosmic rays varies gradually over about two hours. This enhanced anisotropy with a significant amplitude rapidly passed the FOV of the GMDN detectors and was observed as ``bursts'' in some directional channels. This is the first successful result demonstrating that the GMDN is a useful tool for deriving the global cosmic ray anisotropy for timescales shorter than an hour.\par

\begin{figure}[ht!]
\epsscale{1.0}
\plotone{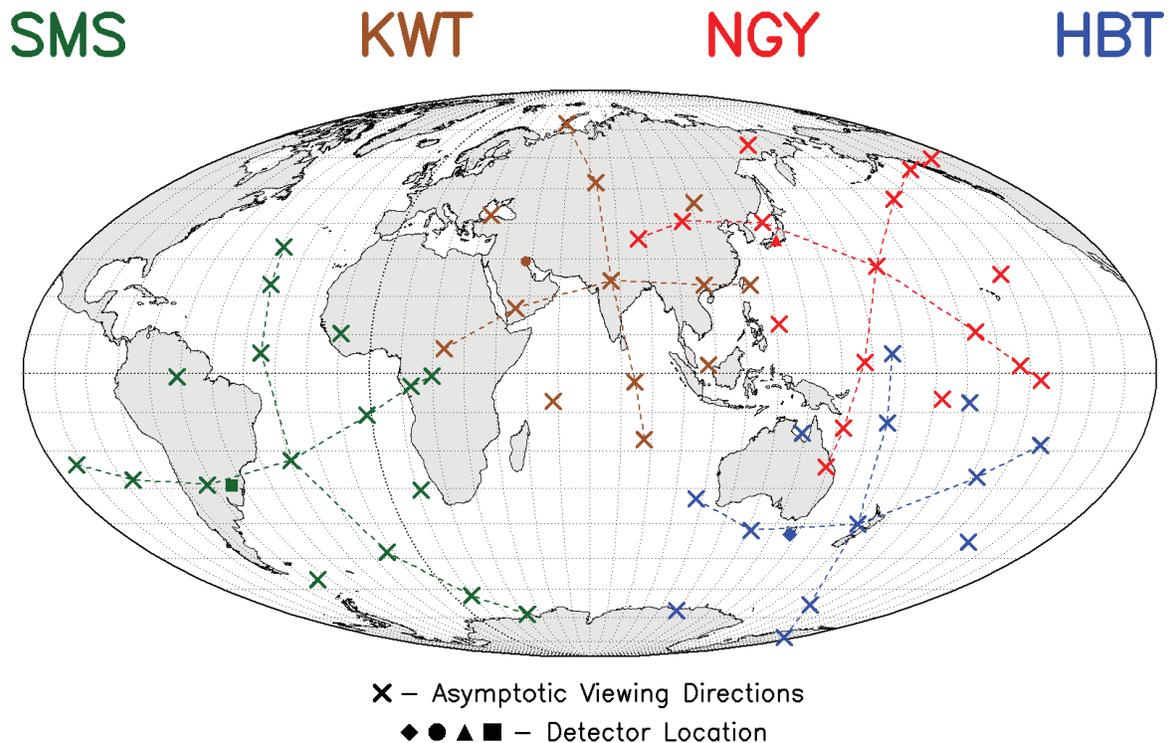}
\vspace{10mm}
\caption{Asymptotic viewing directions of 60 directional channels of the GMDN.
Each colored cross shows the asymptotic direction viewed by a directional channel recording primary cosmic rays with the median primary rigidity $P_m$, while the small solid symbol indicates the location of each detector. Each detector is indicated by different color; Nagoya (NGY) in red, Hobart (HBT) in blue, Kuwait City (KWT) in brown and S\~ao  Martinho da Serra (SMS) in green. Two colored dashed lines for each detector connect the north-south and east-west directional channels, respectively, with the vertical channel at the intersection.
\label{map}}
\end{figure}

\begin{figure}[ht!]
\epsscale{1.0}
\plotone{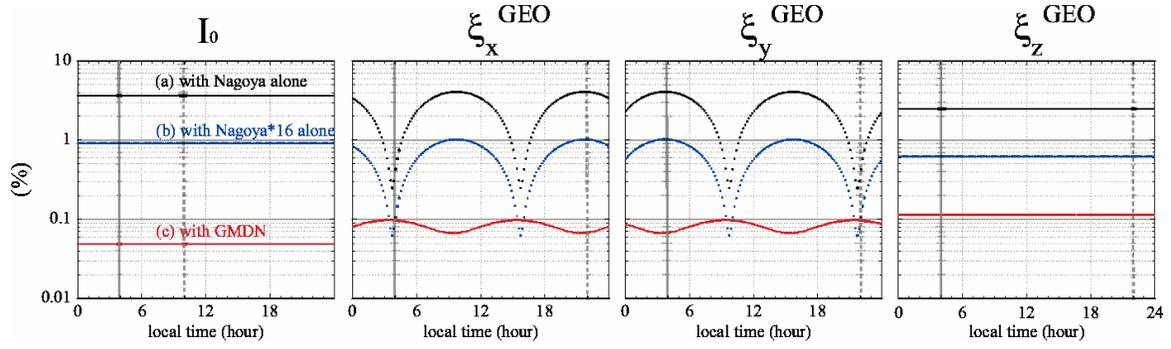}
\vspace{50mm}
\caption{Errors of the best-fit parameters evaluated from the count rate error ($\sigma_{ci,j}$). From left to right, each panel displays the error of $I_0(t)$, $\xi^{\rm GEO}_{x}(t)$, $\xi^{\rm GEO}_{y}(t)$ and $\xi^{\rm GEO}_{z}(t)$ as a function of local time (LT) of Nagoya. Shown in each panel are errors calculated in three cases, (a) best-fitting with only Nagoya data (black curves), (b) best-fitting with only Nagoya data but with count rates virtually enlarged sixteen times mimic the GRAPE-3 (blue curves) and (c) best-fitting with the GMDN data (red curves). The vertical gray solid and dotted lines indicate the LT when the asymptotic viewing direction of Nagoya-V directs along the GEO $y$- and $x$-axes, respectively (see text).
\label{err}}
\end{figure}

By analyzing the muon count rates recorded in nine directional channels of GRAPES-3, Paper 1 concluded that the CRB was unlikely to be caused by a cosmic ray anisotropy outside the magnetosphere. Determining the contribution of the global anisotropy to the observed muon count rate from the observation at a single location on Earth, however, is difficult even for a large detector such as GRAPES-3, particularly when the anisotropy amplitude and orientation rapidly change as indicated in the present paper. The GRAPES-3 is an excellent muon detector with a large detection area (560~m$^{2}$) and muon count rate which is about sixteen times that of Nagoya (36~m$^{2}$), while the width of its FOV is similar to that of Nagoya \citep{Non06}. As seen in Equation~\ref{eq:fit}, the first order anisotropy vector {\boldmath{$\xi$}}$^{\rm GEO}(t)$ produces broad excess and deficit of the relative intensity, each spreading over 180\degr~of longitude. A single muon detector with a limited FOV, therefore, records similar count rates in all directional channels when it monitors the direction parallel or anti-parallel to {\boldmath{$\xi$}}$^{\rm GEO}(t)$, being unable to accurately observe the anisotropy. In order to show this quantitatively, we calculate errors of $I_0(t)$ and {\boldmath{$\xi$}}$^{\rm GEO}(t)$, each as a function of local time (LT) of Nagoya, by propagating the count rate error ($\sigma_{ci,j}$) in Equations \ref{eq:lsm}-\ref{eq:res}.
%In order to show this quantitatively, we performed the following calculations. We first generated 10,000 sample sets of count rates in 60 directional channels at each local time, by adding the statistical fluctuation to each best-fit count rate $I^{fit}_{i,j}(t)$ obtained from $I_0(t)$ and {\boldmath{$\xi$}}$^{\rm GEO}(t)$ at19:45 UT of June 22 when Nagoya-V recorded the highest count rate (see Figure~\ref{obs}(d)). We then performed the best-fit calculation to each sample set by Equations \ref{eq:fit}-\ref{eq:res}, derived the best-fit parameters and calculated the error of each derived parameter from the deviation from that originally used to calculate $I^{fit}_{i,j}(t)$.
Figure~\ref{err} displays errors of $I_0(t)$, $\xi^{\rm GEO}_{x}(t)$, $\xi^{\rm GEO}_{y}(t)$ and $\xi^{\rm GEO}_{z}(t)$ calculated in three cases, (a) best-fitting with only Nagoya data (black curves), (b) best-fitting with only Nagoya data but with the count rates virtually enlarged sixteen times to mimic the GRAPE-3 (blue curves) and (c) best-fitting with the GMDN data (red curves). It is clear from this figure that the error of each parameter in case (a) of best-fitting with Nagoya data alone is much larger than the error in case (c) with the GMDN data. The longitude of asymptotic viewing direction of Nagoya-V is 168.9\degr, while the longitude of the detector's location is 137.0\degr (see Table~\ref{tab1} and Figure~\ref{map}). The center of Nagoya's FOV, therefore, directs toward a certain orientation in space $\sim$2 hours earlier than the corresponding LT at the detector's location and views the direction along the GEO $y$-axis at 90\degr (GEO $x$-axis at 0\degr) GEO longitude at $\sim$04:00 LT (22:00 LT), as indicated by vertical gray solid (dotted) line. It is seen in case (a) that the error of $\xi^{\rm GEO}_{y}(t)$ becomes largest when the FOV of Nagoya directs toward the GEO-$y$ axis, while the error of $\xi^{\rm GEO}_{x}(t)$ becomes smallest at the same local time when Nagoya monitors both the excess and deficit due to $\xi^{\rm GEO}_{x}(t)$ in the FOV. Errors actually become much smaller in case (b) with an enlarged detection area and the reduced $\sigma_{ci,j}$, but still larger than case (c) most of the time. As seen in this figure, errors of $\xi^{\rm GEO}_{x}(t)$ and $\xi^{\rm GEO}_{y}(t)$ in cases (a) and (b) are significantly dependent on the local time, while those in case (c) are much more stable. This is a serious problem for a single detector {\boldmath{$\xi$}}$^{\rm GEO}(t)$ with unknown orientation and amplitude.\par
%Figure~\ref{err} demonstrates that we need a global network of detectors for an accurate observation of the anisotropy.\par

Figure~\ref{err} demonstrates the difficulty of deriving an accurate anisotropy from muon rates in the FOV of a single detector as a function of time. If the anisotropy can be regarded as constant during a day, on the other hand, even a single detector can observe $\xi^{\rm GEO}_{x}$ and $\xi^{\rm GEO}_{y}$ accurately, by measuring the diurnal variations of muon rates in multidirectional channels \citep{Mun14}. In this case, the diurnal variation is generally observed with different phases in different directional channels, because the eastern directional channels observe the anisotropy orientation in space earlier than the western channels according to Earth's spin. If the difference of asymptotic longitudes viewed by the eastern and western channels is $\Delta \phi_{\rm asymp}$, the phase difference of $\Delta \phi_{\rm asymp}/\omega$ (1 hour/15\degr) is expected. Paper 1 concluded that the CRB is not caused by the anisotropy, because no such phase difference is observed in the CRB by GRAPES-3 [see also \citet{Mohanty18}]. Their deduction, however, is valid only when the anisotropy is constant and cannot be applied to a CRB in which the anisotropy rapidly changes in a few hours. It is also noted here that, even in the case of constant anisotropy, a single detector cannot observe $\xi^{\rm GEO}_{z}$ because it produces no diurnal variation.

\begin{figure}[ht!]
\epsscale{1.0}
\plotone{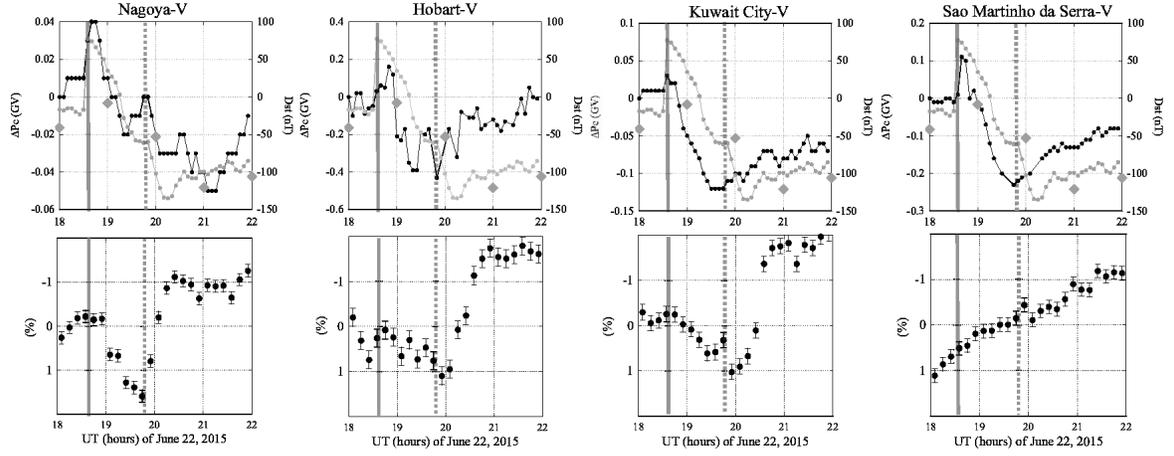}
\vspace{50mm}
\caption{Deviation of the geomagnetic cut-off rigidity from its nominal value calculated by using the latest geomagnetic field model (TS05) during a period between 18:00 and 22:00 UT of June 22. Each upper panel displays the deviation ($\Delta P_{c}$) by a black curve on the left vertical axis calculated as a function of the universal time (UT) for the vertical channel of each detector in GMDN, while each lower panel shows the observed muon count rate with the reversed vertical axis for comparison with $\Delta P_{c}$ in the upper panel. Note different ranges of $\Delta P_{c}$ on left vertical axes in upper panels. Also shown in each upper panel by a gray curve is the reproduced $\it{D_{st}}$ index on the right vertical axis, together with the observed hourly $\it{D_{st}}$ index shown by gray diamonds. Vertical lines in each panel indicate arrival times of the strong shock (gray solid line) and HCS (gray dotted line).
\label{pc}}
\end{figure}

We finally discuss the effect of a major geomagnetic storm recorded in the CRB period. The maximum $\it{K_{p}}$ index of 8+ was recorded during three hours between 18:00 and 21:00 UT of June 22, while the minimum $\it{D_{st}}$ index of -204~nT was recorded at 05:00 UT of June 23\footnote{http://wdc.kugi.kyoto-u.ac.jp/index.html}. Paper 1 concluded that the observed CRB is caused by this geomagnetic storm, because the geomagnetic field is weakened during the storm and the geomagnetic cut-off rigidity ($P_{c}$) of cosmic rays is reduced, allowing more low energy particles to reach the ground level detectors and possibly increasing the muon count rate. To estimate this effect quantitatively, we performed numerical calculations of cosmic ray orbits in the latest model of the geomagnetic field (TS05)\footnote{http://geo.phys.spbu.ru/~tsyganenko/modeling.html}, which is capable of reproducing the geomagnetic field even in a major storm by using the observed solar wind data as inputs \citep{Tsyga05}. In our calculation, we use five-minute averages of these data and reproduce the geomagnetic field every five minutes. This model is also capable of reproducing the variation of $\it{D_{st}}$ index in a good agreement with the observations. Each upper panel of Figure~\ref{pc} shows the reproduced $\it{D_{st}}$ index (gray curve) and the deviation ($\Delta P_{c}$) of $P_{c}$ (black curve) from its nominal value calculated for each vertical channel of GMDN during a period between 18:00 and 22:00 UT on June 22. It is seen that $\Delta P_{c}$ varies roughly in a positive correlation with $\it{D_{st}}$ index and is temporally almost universal and common for the four directional channels, regardless of the location of each detector \citep{Man17}. The amplitude of variation of $\Delta P_{c}$ is, on the other hand, significantly different from one location to the other (see the range of the left vertical axis). By comparing one hour average of $\it{D_{st}}$ index reproduced by the model (gray curve) with the observed hourly $\it{D_{st}}$ index (gray diamond), we verified a good correlation with the correlation coefficient of 0.86 and the regression coefficient of 1.05 together with $\sim$10~\% off-set of the reproduced minimum $\it{D_{st}}$ index. Based on this, we estimate the uncertainty of the reproduced geomagnetic field to be about 10~\%  which is accurate enough for examining the potential effect of the geomagnetic storm on the CRB.\par

\begin{figure}[ht!]
\epsscale{1.0}
\plotone{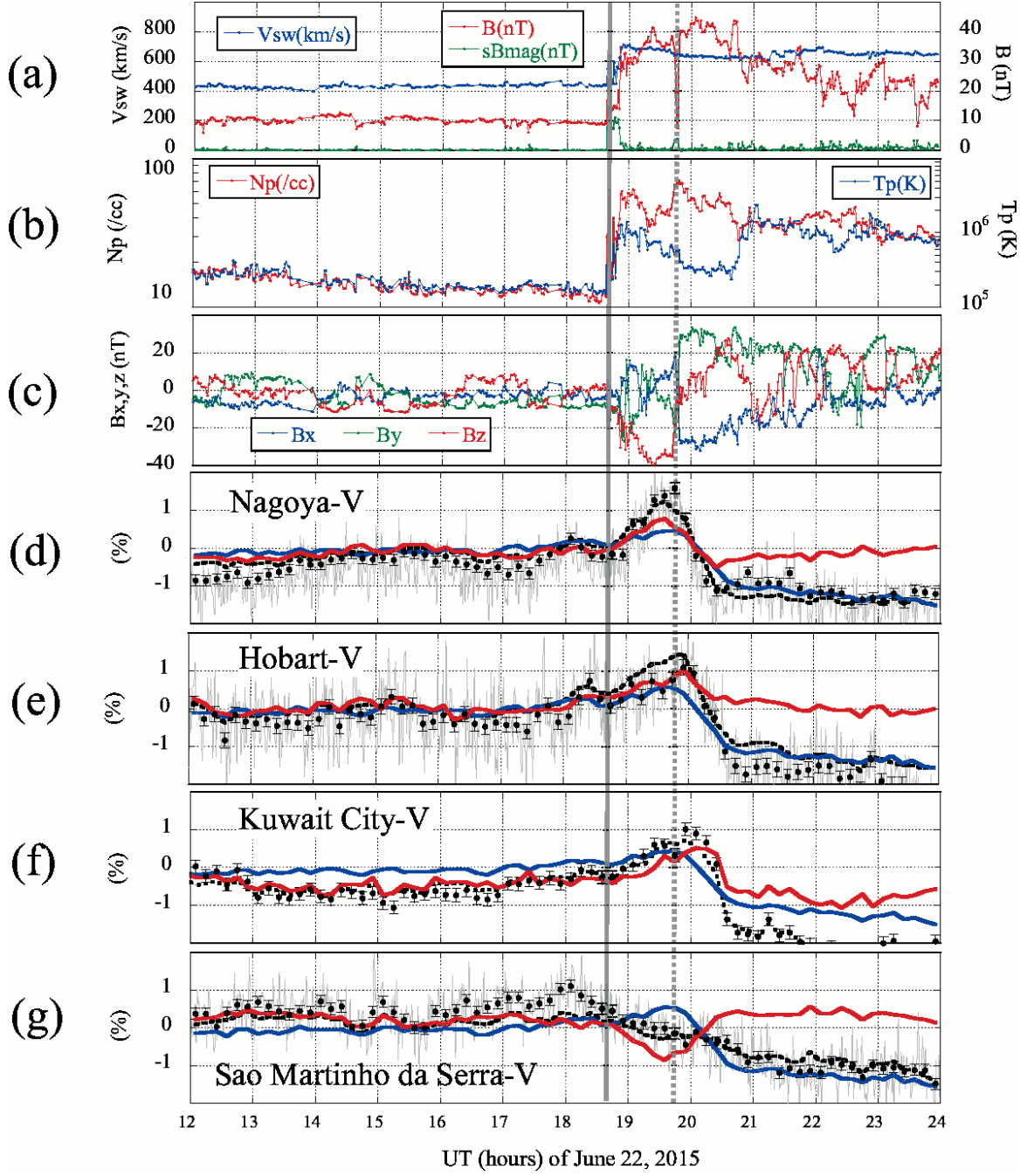}
\caption{Solar wind parameters and muon count rates observed by four vertical channels of the GMDN between 12:00 and 24:00 UT of June 22, 2015.
Figure~\ref{obs}(a)-(c) display 1-minute solar wind data from the OMNIWeb dataset, while solid circles in panels (d)-(g) show $I_{i,j}(t)$ recorded in four vertical channels of the GMDN. Each panel displays (a) solar wind velocity (blue curve) on the left vertical axis and IMF strength (red curve) and its dispersion (green curve) on the right vertical axis, (b) proton density (red curve) and temperature (blue curve) on the left and right vertical axes, respectively, (c) three GSE-components of IMF, (d)-(g) 10-minutes muon count rates recorded in four vertical channels of GMDN (solid circles) each with the count rate error. Panels (a)-(c) show 1-minute data, while gray curves in (d), (e) and (g) also display 1-minute count rate ($\sigma_{ci,j}$). Only 10-minutes data are available from ``Kuwait City'' in (f). The dotted curves in panels (d)-(g) display the best-fit count rate $I^{fit}_{i,j}(t)$, while red and blue curves show contributions from the anisotropy and density to $I^{fit}_{i,j}(t)$ (see text). Vertical gray solid and dotted lines indicate arrival times of the strong shock and HCS, respectively.
\label{obs}}
\end{figure}

\begin{figure}[ht!]
\epsscale{1.0}
\plotone{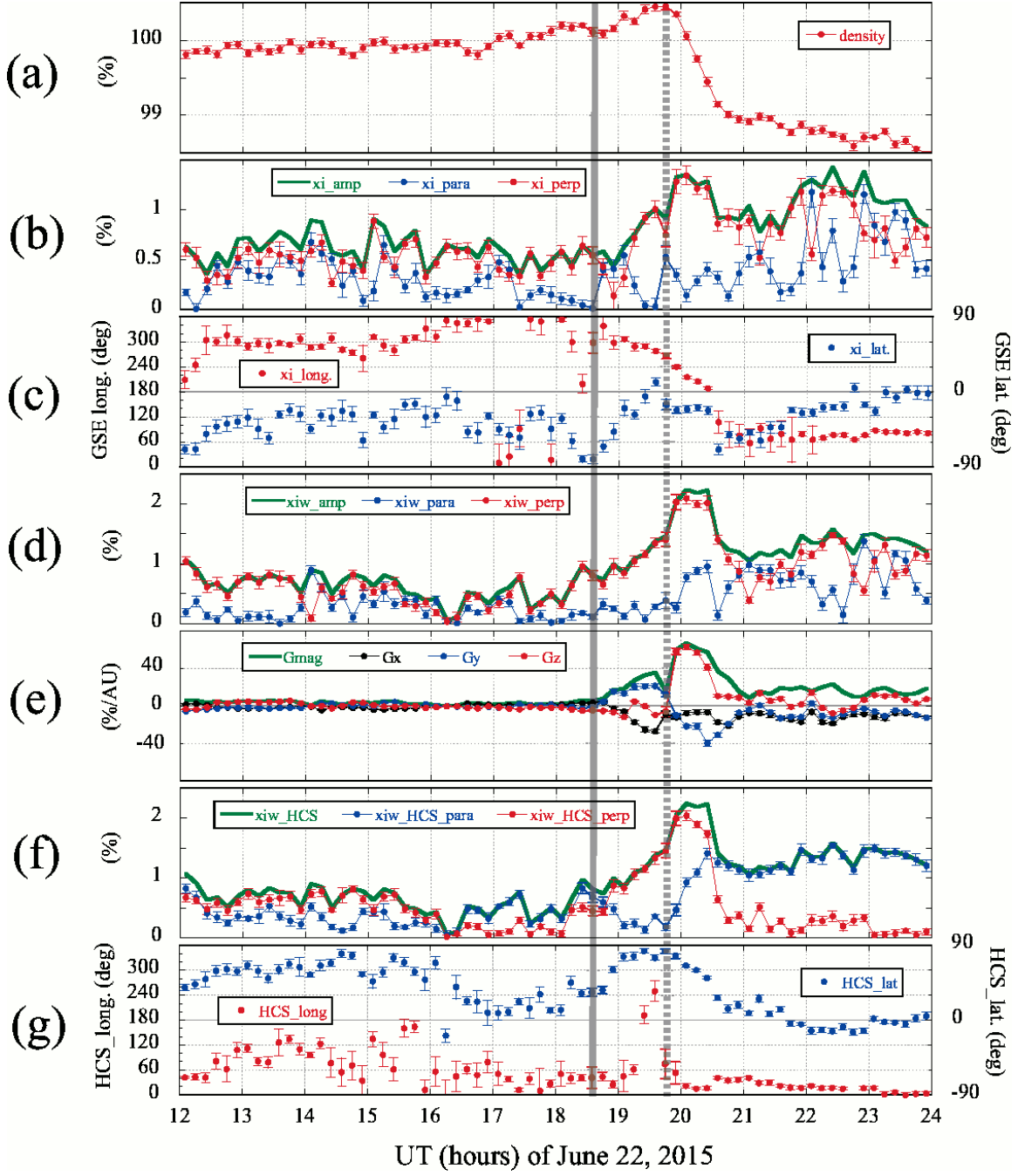}
\caption{Best-fit density and anisotropy for the same period as Figure~\ref{obs}.
Each panel displays best-fit parameters derived from 10-minutes data of GMDN, (a) cosmic ray density, (b) total amplitude of the anisotropy (green curve) and amplitudes of perpendicular (red circles) and parallel (blue circles) components to the local IMF, (c) GSE-longitude (red circles on the left vertical axis) and latitude (blue circles on the right vertical axis) of the anisotropy, (d) amplitude of the total anisotropy corrected for the solar wind convection and Compton-Getting effect (green curve) and amplitudes of perpendicular (red circles) and parallel (blue circles) components to the local IMF, (e) GSE-x (black circles), y (blue circles) and z (red circles) components of the cosmic ray density gradient calculated by assuming the diamagnetic drift streaming for the perpendicular anisotropy, (f) amplitudes of perpendicular (red circles) and parallel (blue circles) components of the anisotropy to the HCS together with the total anisotropy amplitude (green curve), (g) the longitude (red circles) and latitude (blue circles) of the anisotropy in the HCS coordinate system (see Figure~\ref{sch} and text). Vertical lines indicate arrival times of the strong shock (gray solid line) and HCS (gray dotted line).
\label{ani}}
\end{figure}

Listed in the last two columns of Table~\ref{tab1} are the nominal $P_{c}$ and its maximum reduction $\Delta P_{c}$ during a period in Figure~\ref{pc}. By integrating the response function of the atmospheric muon intensity to primary GCRs with respect to the rigidity above $P_{c}$ \citep{Nag71,Mur79,Fuj84}, we calculate the expected increase of muon count rate $\Delta I_{i,j}(t)$ as listed in Table ~\ref{tab1}. The expected $\Delta I_{i,j}(t)$ is only $\sim$0.3~\% at most and much smaller than the observed amplitude of the CRB in Figure~\ref{obs}. The reduction of $P_{c}$ is -0.64~GV and second largest for Hobart-V, but $\Delta I_{i,j}(t)$ is almost zero. This is because $P_{c}$ for Hobart-V is already lower than the atmospheric threshold rigidity for producing muons and the reduction of $P_{c}$ causes no further increase in the muon count rate. Contrary to this, the CRB is recorded also in Hobart-V as seen in Figure~\ref{obs}(e). The largest $\Delta I_{i,j}(t)$ would be expected in S\~ao  Martinho da Serra-V, but no CRB is observed in this channel, as shown in Figure~\ref{obs}(g). The observed muon count rate in the lower panel of Figure~\ref{pc} actually shows no clear correlation with $\Delta P_{c}$ in the upper panel. We also confirmed that the asymptotic viewing direction during the storm varies only a few degrees at most, insufficient to affect the directional response of muon detectors to primary cosmic ray intensity and result in the observed CRB. Based on these results, we conclude that the CRB is not caused by the geomagnetic storm.\par

\acknowledgments
This work is supported in part by the joint research programs of the Institute for Space-Earth Environmental Research (ISEE), Nagoya University and the Institute for Cosmic Ray Research (ICRR), University of Tokyo. The observations are supported by Nagoya University with the Nagoya muon detector, by CAPES, INPE, and UFSM with the S\~ao Martinho da Serra muon detector, and by the Australian Antarctic Division with the Hobart muon detector. The observation with Kuwait City muon detector is supported by the project SP01/09 of the Research Administration of Kuwait University. CRB thanks to Sao Paulo Research Fundation (grant 14/24711-6). RRSM thanks CNPq (grant 152050/2016-7) and the China-Brazil Joint Laboratory for Space Weather. JK was partly supported by the Joint Research Program of the Institute for Space Earth Environmental Research, Nagoya University, Japan. Authors thank Dr. Ryuho Kataoka of the National Institute of Polar Research in Japan for providing us with the information of Tsyaganeno's geomagnetic field models. They also thank Dr. Yoji Hasegawa of Shinshu University for his help in performing MC simulations by GEANT4. The OMNIWeb dataset of the solar wind and IMF parameters is provided by the Goddard Space Flight Center, NASA, USA, while the MHD simulation of the space weather during the CRB is provided by the WSA-Enlil solar wind prediction, National Centers of Environmental Information, NOAA, USA. The hourly $\it{D_{st}}$ index is provided by the WDC for Geomagnetism, Kyoto, Japan. We are grateful to the anonymous referee for thoughtful comments which were useful for developing a clearer presentation of this paper.

\newpage




\newpage

\begin{deluxetable}{ccccccccc}
\tabletypesize{\scriptsize}
\tablewidth{0pt}
\tablenum{1}
\tablecaption{Characteristics and estimated responses to the geomagnetic storm of four vertical channels of GMDN$^{a}$.\label{tab1}}
\tablehead{
\colhead{detector name} & \colhead{detector type} & \colhead{count rate} & \colhead{$\sigma_{ci,j}$} & \colhead{$P_{m}$} & \colhead{$\phi_{\rm asymp}$} & \colhead{$\lambda_{\rm asymp}$} & \colhead{$P_{c}~\&~\Delta P_{c}$} & \colhead{$\Delta I_{i,j}(t)$}  \\
\colhead{(no. of directions)} & \colhead{(detection area)} & \colhead{$(10^{4}$/10-min.)} & \colhead{(\%)} & \colhead{(GV)} & \colhead{(\degr)} & \colhead{(\degr)} & \colhead{(GV)} & \colhead{(\%)}
}
\startdata
Nagoya-V&PS&47.8&0.14&58.4&168.9&27.7&12.3 \& -0.07&+0.08\\
(17)&(36 m$^{2}$)&(3.0$\sim$47.8)&(0.14$\sim$0.58)&(58.4$\sim$106.9)&(89.1$\sim$235.8)&(-24.4$\sim$64.0)&&\\
Hobart-V&PS&23.7&0.21&53.1&171.0&-39.8&1.6 \& -0.64&+0.00\\
(13)&(16 m$^{2}$)&(3.2$\sim$23.7)&(0.21$\sim$0.56)&(53.1$\sim$74.0)&(108.0$\sim$237.1)&(-76.7$\sim$5.0)&&\\
Kuwait City-V&PCT&24.2&0.20&59.8&78.4&23.8&13.0 \& -0.10&+0.08\\
(13)&(18.5 m$^{2}$)&(4.0$\sim$24.2)&(0.20$\sim$0.50)&(59.8$\sim$94.5)&(24.6$\sim$127.9)&(-19.2$\sim$72.0)&&\\
S\~ao  Martinho da Serra-V&PS&42.3&0.15&54.3&331.4&-22.4&9.8 \& -0.82&+0.29\\
(17)&(32.0 m$^{2}$)&(0.7$\sim$42.3)&(0.15$\sim$1.22)&(54.3$\sim$98.4)&(259.4$\sim$32.7)&(-67.1$\sim$33.4)&&\\
\enddata

\tablenotetext{a}{Following the detector name (number of directional channels) and type (detection area), the average 10-minutes count rate, count rate error ($\sigma_{ci,j}$), median primary rigidity ($P_m$), geographic longitude ($\phi_{\rm asymp}$) and latitude ($\lambda_{\rm asymp}$) of asymptotic viewing direction outside the magnetosphere are listed for each of four vertical channels of the GMDN. Each number in brackets in columns 3-7 indicates a range of corresponding parameter covered by all directional channels available in each detector. The last two columns are the average geomagnetic cut-off rigidity ($P_c$), its maximum reduction ($\Delta P_{c}$) and the maximum increase of count rate expected from $\Delta P_{c}$, which are calculated using the model geomagnetic field during the geomagnetic storm on June 22, 2015 and the response function of each directional channel to primary GCRs (see text).}

\end{deluxetable}
%\vspace{5mm}

%\newpage
\clearpage

%% The "ht!" tells LaTeX to put the figure "here" first, at the "top" next
%% and to override the normal way of calculating a float position

\newpage

\newpage

\newpage

\newpage

\newpage

\newpage

\end{document}